\documentclass[aps,pre,twocolumn,superscriptaddress,nofootinbib]{revtex4-2}

\usepackage{graphicx}
\usepackage{amsmath,amssymb}
\usepackage{xcolor}

\usepackage{booktabs}
\usepackage{nameref}
\usepackage{hyperref}
\usepackage[utf8]{inputenc}
\usepackage{newunicodechar}
\newunicodechar{₂}{$_2$}

\begin{document}

\title{A Nonhomogeneous Porous-Medium Equation for Field Scale CO$_2$ Plume Spreading}

\author{Fernando Alonso-Marroqu\'in}
\email{fernando@quantumfi.net}
\affiliation{Department of Computational Physics for Engineering Materials, ETH Zurich, 8092 Zurich, Switzerland}

\author{Christian Tantardini}
\email{christiantantardini@ymail.com}
\affiliation{Center for Integrative Petroleum Research, King Fahd University of Petroleum and Minerals, Dhahran 31261, Saudi Arabia}

\begin{abstract}
We derive a nonlinear diffusion model for field scale CO$_2$ plume spreading from a Global Buckley--Leverett component balance. The reduced variable $u$ is the vertically averaged mobile  gas phase CO$_2$ content normalized by its maximum column value; under vertical segregation, $u=h/H$, where $h$ is plume thickness and $H$ is aquifer thickness. The resulting equation is a nonhomogeneous porous medium type equation in which nonlinear lateral spreading is coupled to source/sink terms for injection, dissolution, mineral fixation, and retention. Using the nonlinear diffusivity $D_u(u)\simeq D_0u^{1-q}$, we analyze Barenblatt-type profiles with prescribed mobile mass and a capped plume constrained by $0\le u\le1$. The capped solution contains a ful-thickness core of radius $a(t)$ and a compact plume edge $R(t)$. Constant net mobile injection can sustain the core and gives square-root growth of $R(t)$, whereas shut-in or weak mobile addition causes the core to shrink and disappear. We compare these regimes with equivalent radii from time lapse seismic plume maps at Sleipner, Aquistore, and Weyburn--Midale. The data distinguish injection controlled growth, delayed layer filling, and tail dominated redistribution, but do not determine a unique nonlinear exponent. The model provides an analytical reference for interpreting plume footprint evolution while separating cumulative injected CO$_2$ from mobile  gas phase CO$_2$.
\end{abstract}

\date{\today}

\maketitle

\section{Introduction}

The spreading of a confined buoyant fluid in a porous medium is a classical nonlinear free boundary problem \cite{barenblatt1996scaling,Pattle1959,Vazquez2007PME}. Its reduced transport structure is also connected to nonlinear Fokker--Planck equations with exact time-dependent \(q\)-Gaussian solutions \cite{TsallisBukman1996}. The same nonlinear diffusion language has been used to describe \(q\)-Gaussian spreading in stock-market fluctuations \cite{alonso2019q} and porous medium dynamics with variable order in price return data \cite{Tang2024variable,Tang2025bitcoin}. Here, the nonlinear exponent is used for a physical transport problem: the field scale spreading of mobile  gas phase CO$_2$ in a confined aquifer.

In geological CO$_2$ storage, plume migration is commonly modeled using vertically integrated Darcy equations or sharp interface gravity current reductions \cite{Nordbotten2005,NordbottenCelia2011,HuppertWoods1995,Hesse2008}. These models reduce the vertical saturation structure to a plume thickness variable, use the CO$_2$-brine density contrast to derive buoyancy driven spreading laws, and predict the evolution of plume pressure and radial extent during injection and after shut-in. They provide the reduced theory reference for interpreting observed plume footprints, but they also show why plume radius cannot be inferred from injected mass alone.

This point matters because several physical processes modify the relation between injected CO$_2$ and mobile plume extent. Analytical models of CO$_2$ injection into saline aquifers show that, during active injection, near well displacement is primarily controlled by the pressure gradient Darcy flux generated by injection overpressure \cite{Nordbotten2005}. Buoyancy acts from the beginning by segregating the CO$_2$ from brine into a gas phase and forming a plume beneath the caprock. Once this segregated layer is established, gravity also contributes to horizontal plume migration through the porous layer gravity current mechanism, especially farther from the injector, beneath dipping seals, or after pressure relaxation \cite{HuppertWoods1995,MacMinn2010}. Capillary forces act differently: they control pore entry, capillary transition zones, residual brine saturation within the CO$_2$-rich region, and residual trapping of disconnected CO$_2$ during imbibition \cite{Leverett1941,Juanes2006,MacMinn2010}. Dissolution and mineral fixation further remove CO$_2$ from the mobile gas phase and reduce the amount available for continued horizontal spreading \cite{MacMinn2011}. These mechanisms motivate a reduced plume model that distinguishes the mobile  gas phase CO$_2$ amount controlling the plume footprint from the local processes that modify saturation, trapping, and phase exchange.

The reduced model developed here therefore focuses on the large scale
pressure--gravity spreading of the mobile plume. Capillary trapping,
dissolution, and hysteretic relative permeability are not resolved as separate mechanisms; instead, their net effect is absorbed into the interpretation of the scalar mobile plume variable. This choice also motivates a different route from the standard gravity current derivation. Instead of postulating a vertically
equilibrated sharp interface model as the starting point, we begin from the Global Buckley--Leverett (GBL) formulation for multicomponent, multiphase flow
\cite{TantardiniAlonsoMarroquin2026GBL}. 
In that conservative formulation, the natural transported quantity is the amount of a chemical component, rather than the density of a fluid phases. This suggests a scalar plume variable based on the
mobile free phase CO$_2$ content. Under vertical segregation, this content is proportional to the vertically averaged thickness of the connected gas  layer normalized by the aquifer thickness. The variable is therefore intrinsically bounded: the mobile layer cannot exceed the formation thickness.

Starting from this scalar plume variable, we derive a reduced lateral spreading model by combining mass balance with a coarse-grained relation for horizontal transport. At that stage, the effective local nonlinear diffusivity is composed of three contributions, representing pressure redistribution, buoyancy, and capillarity. The result is a nonlinear diffusion equation for the mobile plume thickness. The corresponding nonlinearity exponent is not interpreted as a microscopic material constant; rather, it is an effective large scale parameter summarizing the Darcy scale spreading response after vertical reduction.

The finite aquifer thickness changes the admissible similarity structure relative to the classical Barenblatt--Pattle profiles. In the porous-medium equation, these profiles have a compact outer front, but their central amplitude is not constrained by a prescribed layer thickness \cite{Pattle1959,barenblatt1996scaling}. A CO$_2$ plume in a confined aquifer is different: the mobile gas  layer cannot become thicker than the formation. If an unconstrained Barenblatt-type profile would exceed this thickness near the injection region, the excess cannot be represented by increasing the local amplitude. Instead, the plume develops a ful-thickness inner core connected to an outer nonlinear diffusion tail. The resulting reduced problem has two moving radii: the radius of the ful-thickness core and the outer edge of the mobile plume.

The aim of this paper is to derive and analyze this capped Barenblatt-type construction in a form consistent with the GBL parent description. We identify the similarity structure during constant injection and after shut-in, show how the finite thickness constraint modifies the standard source driven and source free porous-medium scalings, and obtain explicit relations for the evolution of the core radius and the plume edge. Published seismic footprint data are used only as an external consistency check on the resulting scaling ranges, not as an input to the derivation.

\section{Model}
\label{sec:model}

\begin{figure}[t]
    \centering
    \includegraphics[width=\linewidth,trim={4cm 0.5cm 0cm 1cm},clip]{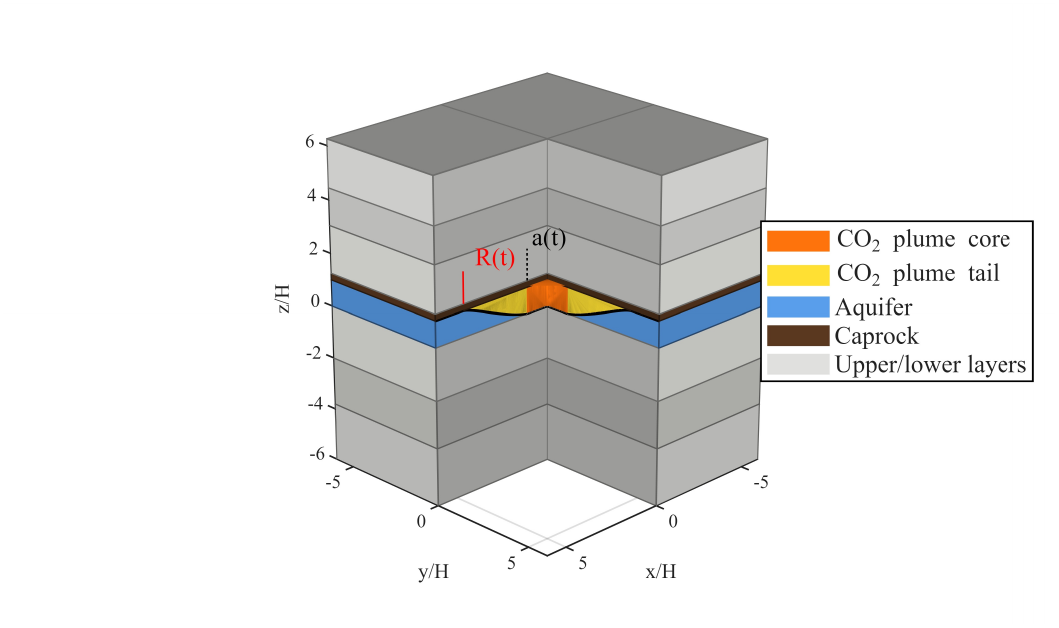}
    \caption{Geometrical description of the axisymmetric CO$_2$ plume model. The orange region denotes the plume core, $0\le r\le a(t)$, where CO$_2$ occupies the full aquifer thickness. The yellow region denotes the plume tail, $a(t)<r\le R(t)$, where the plume thickness decreases toward the outer radius $R(t)$. The blue layer is the aquifer, the brown layer is the caprock, and the gray layers indicate the surrounding stratigraphy.}
    \label{fig:plume_geometry}
\end{figure}

We consider the lateral spreading of a CO$_2$ plume in a horizontal aquifer of thickness $H$, bounded above by a caprock. The plume is represented by an axisymmetric gravity current geometry in which the vertical structure is reduced to the dimensionless radial field $u(r,t)$. The variable $u(r,t)$ measures the local mobile  gas phase CO$_2$ column content normalized by its ful-thickness value. Under vertical segregation, this reduces to $u(r,t)=h(r,t)/H$, where $h(r,t)$ is the CO$_2$-occupied thickness. The geometry separates the plume into two radial regions. The inner region $0\le r\le a(t)$ is the plume core, where the CO$_2$-rich layer occupies the full aquifer thickness and $u(r,t)=1$. The outer region $a(t)<r\le R(t)$ is the plume tail, where $u(r,t)$ decreases continuously from one to zero, with $u(R(t),t)=0$.
Figure~\ref{fig:plume_geometry} summarizes the axisymmetric geometry and fixes the notation used throughout the model. The plume is described by the bounded scalar field $u(r,t)$, which represents the vertically reduced mobile  gas phase CO$_2$ content. The inner radius $a(t)$ marks the ful-thickness plume core, where $u=1$, and the outer radius $R(t)$ marks the compact edge of the mobile plume, where $u=0$. This geometrical description provides the link between the vertically averaged transport equation and the later Barenblatt-type radius laws.

The model is then developed in four steps. Section~\ref{sec:gbl_to_scalar} starts from the component molar CO$_2$ balance in the Global Buckley--Leverett formulation and reduces it to the bounded plume scale variable $u$. Section~\ref{sec:gbl_pme} projects the Darcy/GBL current onto this scalar variable, giving a nonhomogeneous porous-medium equation with effective transport coefficient $D_u=D_p+D_\mathrm{grav}+D_c$, where $D_p$, $D_\mathrm{grav}$, and $D_c$ account for transport driven by pressure, gravity, and capillarity, respectively. The same equation retains the net source/sink term $\gamma^{2D}/n_\ast$, which accounts for CO$_2$ injection and for removal of mobile  gas phase CO$_2$ by dissolution, mineral fixation, and retention. Section~\ref{sec:barenblatt} isolates the spreading structure of the nonlinear operator through a mass-controlled Barenblatt reference profile, in which the source/sink history enters through the mobile inventory $M_u(t)$. Finally, Section~\ref{sec:capped_barenblatt} enforces the physical bound $0\le u\le1$ by introducing a capped Barenblatt-type plume with a ful-thickness core of radius $a(t)$ and an outer plume radius $R(t)$.

\subsection{From Global Buckley--Leverett transport to a plume scale scalar variable}
\label{sec:gbl_to_scalar}

This section derives the scalar field governing the reduced nonlinear spreading description of the plume. Beginning with the conservative component balance of the Global Buckley--Leverett (GBL) formulation, we develop a sequence of plume scale reductions that yields a bounded scalar variable and an effective lateral transport law \cite{TantardiniAlonsoMarroquin2026GBL}

For the present reduced theory we write the CO$_2$ balance in component molar
form. This is the natural choice because the conserved object is not a fluid
phase, but the CO$_2$ component. During storage, the same component may be
present in a gas  phase, dissolved in the brine (i.e., water phase), or associated
with the rock through retention, adsorption, or mineral reactions. These
contributions should not be combined by adding phase densities directly,
because gas, brine, and rock associated storage are different carriers. They
can, however, be combined as amounts of the same chemical component. We
therefore use the total CO$_2$ molar concentration per bulk volume as the
parent conserved variable, with molar fractions specifying the CO$_2$
composition of each fluid phase.
The passage from the full conservative balance to the scalar plume equation
then involves three steps: (i) isolation of the plume  gas phase CO$_2$
contribution, (ii) vertical equilibrium reduction to a bounded plume scale
variable, and (iii) closure of the projected horizontal molar current.

We distinguish phase quantities from component quantities. The
symbols \(n_g\) and \(n_w\) denote the total molar concentrations of the
gas and water (i.e. brine) phases, respectively. By contrast,
\(n_{\mathrm{CO_2}}^{g}\), \(n_{\mathrm{CO_2}}^{w}\), and
\(n_{\mathrm{CO_2}}^{r}\) denote CO$_2$ molar concentrations per bulk volume
in the gas phase, in the water phase, and in the rock (i.e., solid) associated
contribution. The molar fraction of CO$_2$ in phase \(\alpha\) is denoted by
\(\chi_{\mathrm{CO_2},\alpha}\), with \(\alpha=g\) for the gas phase and
\(\alpha=w\) for the water phase.

The single continuum component molar balance for CO$_2$ can be written as
\begin{widetext}
\begin{equation}
\frac{\partial}{\partial t}
\left[
\phi
\sum_{\alpha=g,w}
S_\alpha n_\alpha \chi_{\mathrm{CO_2},\alpha}
+
n_{\mathrm{CO_2}}^{r}
\right]
+
\nabla\!\cdot
\left[
\sum_{\alpha=g,w}
n_\alpha \chi_{\mathrm{CO_2},\alpha}\mathbf v_\alpha
-
\sum_{\alpha=g,w}
\mathbf j_{\mathrm{CO_2},\alpha}^{\mathrm{na}}
\right]
=
\gamma_{\mathrm{CO_2}} .
\label{eq:gbl_molar_component_balance_plume}
\end{equation}
\end{widetext}
Here \(\phi\) is porosity, \(S_\alpha\) is phase saturation,
\(\mathbf v_\alpha\) is the phase Darcy velocity, and
\(\gamma_{\mathrm{CO_2}}\) is a local CO$_2$ molar source per bulk volume and time.
The product \(n_\alpha\chi_{\mathrm{CO_2},\alpha}\) is the CO$_2$ molar
concentration within phase \(\alpha\), while
\(\phi S_\alpha n_\alpha\chi_{\mathrm{CO_2},\alpha}\) is the corresponding
CO$_2$ molar concentration per bulk volume. The vector
\(\mathbf j_{\mathrm{CO_2},\alpha}^{\mathrm{na}}\) denotes the nonadvective
CO$_2$ molar current density in phase \(\alpha\). The rock associated term
\(n_{\mathrm{CO_2}}^{r}\) is a CO$_2$ molar concentration per bulk volume,
not a molar fraction, because the rock is not treated as a mobile fluid phase
with a phase composition.

The total CO$_2$ molar concentration per bulk volume is decomposed as
\begin{equation}
n_{\mathrm{CO_2}}^{\mathrm{tot}}
=
n_{\mathrm{CO_2}}^{g}
+
n_{\mathrm{CO_2}}^{w}
+
n_{\mathrm{CO_2}}^{r}.
\label{eq:nCO2_total_decomposition}
\end{equation}
The first two terms are
\begin{equation}
n_{\mathrm{CO_2}}^{g}
=
\phi S_g n_g\chi_{\mathrm{CO_2},g},
\label{eq:nCO2_gas_def}
\end{equation}
and
\begin{equation}
n_{\mathrm{CO_2}}^{w}
=
\phi S_w n_w\chi_{\mathrm{CO_2},w}.
\label{eq:nCO2_aqueous_def}
\end{equation}
Thus \(n_{\mathrm{CO_2}}^{g}\) is the CO$_2$ contained in the gas phase
per bulk volume, whereas \(n_{\mathrm{CO_2}}^{w}\) is the dissolved CO$_2$
contained in the  water phase per bulk volume. The rock associated
contribution \(n_{\mathrm{CO_2}}^{r}\) represents CO$_2$ retained, adsorbed,
or mineral fixed by the solid matrix. It may be prescribed by a constitutive
rock storage model, constrained by solid phase carbon or mineral carbonate
measurements, or inferred by closure when
\(n_{\mathrm{CO_2}}^{\mathrm{tot}}\), \(n_{\mathrm{CO_2}}^{g}\), and
\(n_{\mathrm{CO_2}}^{w}\) are known:
\begin{equation}
n_{\mathrm{CO_2}}^{r}
=
n_{\mathrm{CO_2}}^{\mathrm{tot}}
-
n_{\mathrm{CO_2}}^{g}
-
n_{\mathrm{CO_2}}^{w}.
\label{eq:nCO2_rock_by_closure}
\end{equation}
This notation keeps the phase molar concentrations \(n_g,n_w\) distinct from
the CO$_2$ component concentrations
\(n_{\mathrm{CO_2}}^{g}\), \(n_{\mathrm{CO_2}}^{w}\), and
\(n_{\mathrm{CO_2}}^{r}\). The conserved quantity
\(n_{\mathrm{CO_2}}^{\mathrm{tot}}\) is therefore the total CO$_2$ molar
concentration per bulk volume, not the total gas molar concentration.

The corresponding advective CO$_2$ molar current densities are
\begin{equation}
\mathbf j_{\mathrm{CO_2}}^{g}
=
n_g\chi_{\mathrm{CO_2},g}\mathbf v_g
\label{eq:jCO2_gas_def}
\end{equation}
for the gas phase, and
\begin{equation}
\mathbf j_{\mathrm{CO_2}}^{w}
=
n_w\chi_{\mathrm{CO_2},w}\mathbf v_w
\label{eq:jCO2_aqueous_def}
\end{equation}
for the  water phase. The total nonadvective CO$_2$ molar current density is
\begin{equation}
\mathbf j_{\mathrm{CO_2}}^{\mathrm{na}}
=
\sum_{\alpha=g,w}
\mathbf j_{\mathrm{CO_2},\alpha}^{\mathrm{na}} .
\label{eq:jCO2_na_def}
\end{equation}

The decomposition in Eqs.~\eqref{eq:jCO2_gas_def}--\eqref{eq:jCO2_na_def}
is a carrier decomposition of the CO$_2$ current. It separates the CO$_2$
transport according to where the component is carried: in the gas phase, in the
aqueous phase, or by nonadvective transport. This decomposition is therefore
not yet a decomposition by driving force. The latter is introduced only after
the mobile gas-phase contribution has been isolated and closed by Darcy's law.
In the reduced plume model below, the lateral footprint is controlled by the
connected gas-phase CO$_2$ current, whereas dissolved CO$_2$, rock-associated
storage, and other removal mechanisms enter the mobile-gas balance through the
net source/sink term.

With these definitions, the conservative parent balance can be written as
\begin{equation}
\frac{\partial n_{\mathrm{CO_2}}^{\mathrm{tot}}}{\partial t}
+
\nabla\!\cdot
\left(
\mathbf j_{\mathrm{CO_2}}^{g}
+
\mathbf j_{\mathrm{CO_2}}^{w}
-
\mathbf j_{\mathrm{CO_2}}^{\mathrm{na}}
\right)
=
\gamma_{\mathrm{CO_2}} .
\label{eq:nCO2_total_balance}
\end{equation}

Eq.~\eqref{eq:nCO2_total_balance} keeps the complete CO$_2$ molar concentration per bulk volume. The reduced plume equation is obtained by isolating the mobile free-phase CO$_2$ contribution \(n_{\mathrm{CO_2}}^{g}\), because this is the contribution that defines the mobile plume footprint and, after vertical averaging, the bounded plume-scale variable. The aqueous, rock-associated, and nonadvective contributions are not discarded; instead, they are absorbed into an effective source term for the mobile free phase.

Using the decomposition in Eq.~\eqref{eq:nCO2_total_decomposition}, Eq.~\eqref{eq:nCO2_total_balance} can be rearranged as
\begin{equation}
\frac{\partial n_{\mathrm{CO_2}}^{g}}{\partial t}
+
\nabla\!\cdot \mathbf j_{\mathrm{CO_2}}^{g}
=
\gamma_g ,
\label{eq:nCO2_gas_balance_general}
\end{equation}
where the effective mobile free-phase source is
\begin{equation}
\begin{aligned}
\gamma_g
&=\,
\gamma_{\mathrm{CO_2}}
-
\frac{\partial}{\partial t}
\left(
n_{\mathrm{CO_2}}^{w}
+
n_{\mathrm{CO_2}}^{r}
\right)
\\
&-
\nabla\!\cdot
\left(
\mathbf j_{\mathrm{CO_2}}^{w}
-
\mathbf j_{\mathrm{CO_2}}^{\mathrm{na}}
\right).
\end{aligned}
\label{eq:qg_def}
\end{equation}
Thus \(\gamma_g\) is not the total injected CO$_2$ source. It is the net molar source of mobile free-phase CO$_2$ after accounting for dissolution into brine, rock-associated storage, and nonadvective CO$_2$ redistribution. When aqueous transport and nonadvective redistribution are negligible or absorbed into local exchange rates, Eq.~\eqref{eq:qg_def} reduces to
\begin{equation}
\gamma_g
=
\gamma_{\mathrm{CO_2}}
-
r_{\mathrm{diss}}
-
r_{\mathrm{rock}} .
\label{eq:qg_simplified_exchange}
\end{equation}
Here \(r_{\mathrm{diss}}\) and \(r_{\mathrm{rock}}\) denote the effective rates at which CO$_2$ leaves the mobile free phase by dissolution and rock-associated storage.

To close Eq.~\eqref{eq:nCO2_gas_balance_general}, we express the  gas phase
CO$_2$ molar current density in terms of the Darcy velocity of the gas 
phase. In the reduced plume model, the connected gas  phase is treated as
pure CO$_2$, so that \(\chi_{\mathrm{CO_2},g}=1\).
With this approximation, Eq.~\eqref{eq:jCO2_gas_def} becomes
\begin{equation}
\mathbf j_{\mathrm{CO_2}}^{g}
=
n_g\mathbf v_g ,
\label{eq:jCO2_gas_pure}
\end{equation}
and Eq.~\eqref{eq:nCO2_gas_def} reduces to
\begin{equation}
n_{\mathrm{CO_2}}^{g}
=
\phi S_g n_g .
\label{eq:nCO2_gas_pure}
\end{equation}

Thus, within the connected gas phase, the phase molar concentration \(n_g\) also equals the CO$_2$ molar concentration in that phase. We first close the local Darcy-scale current and only afterwards apply vertical averaging and projection onto the bounded plume variable \(u\).

The local Darcy laws for the gas and water phases are
\begin{align}
\mathbf v_g
&=
-k\lambda_g\left(\nabla p_g-\rho_g\mathbf g\right),
\qquad
\lambda_g=\frac{k_{rg}}{\mu_g},
\label{eq:darcy_gas_phase_closure}
\\
\mathbf v_w
&=
-k\lambda_w\left(\nabla p_w-\rho_w\mathbf g\right),
\,\,\quad
\lambda_w=\frac{k_{rw}}{\mu_w}.
\label{eq:darcy_water_phase_closure}
\end{align}
The total Darcy velocity is \(\mathbf v_t=\mathbf v_g+\mathbf v_w\). Here \(k\) is intrinsic permeability, \(\lambda_\alpha\) is phase mobility, \(p_\alpha\) is phase pressure, \(\rho_\alpha\) is phase density entering the body-force term, and \(\mathbf g\) is the gravitational acceleration vector. With \(f_g=\lambda_g/\Lambda_t\), \(\Lambda_t=\lambda_g+\lambda_w\), \(p_c=p_g-p_w\), and \(\Delta\rho=\rho_w-\rho_g\), Eqs.~\eqref{eq:darcy_gas_phase_closure} and \eqref{eq:darcy_water_phase_closure} give the fractional-flow form
\begin{equation}
\mathbf v_g=f_g\mathbf v_t-k\frac{\lambda_g\lambda_w}{\Lambda_t}\left(\nabla p_c+\Delta\rho\,\mathbf g\right).
\label{eq:vg_fractional_flow_closure}
\end{equation}
Equation~\eqref{eq:vg_fractional_flow_closure} follows by eliminating \(\mathbf v_w\) from \(\mathbf v_t=\mathbf v_g+\mathbf v_w\). The sign of the gravity term follows from the definitions \(p_c=p_g-p_w\) and \(\Delta\rho=\rho_w-\rho_g\).

Since the connected gas phase is treated as nearly pure CO$_2$, 
Eq.~\eqref{eq:jCO2_gas_def} gives
\(\mathbf j_{\mathrm{CO_2}}^g=n_g\mathbf v_g\). Multiplying
Eq.~\eqref{eq:vg_fractional_flow_closure} by \(n_g\) gives
\begin{equation}
\begin{aligned}
\mathbf j_{\mathrm{CO_2}}^g
&=
\underbrace{n_g f_g\mathbf v_t}_{\mathbf j_{\mathrm{tf}}}
\underbrace{
-n_g k\frac{\lambda_g\lambda_w}{\Lambda_t}\nabla p_c
}_{\mathbf j_c}
\underbrace{
-n_g k\frac{\lambda_g\lambda_w}{\Lambda_t}\Delta\rho\,\mathbf g
}_{\mathbf j_{\mathrm{grav}}}.
\end{aligned}
\label{eq:jCO2_gas_mechanism_split}
\end{equation}
This is a mechanism-level decomposition of the already isolated mobile
gas-phase CO$_2$ current. It is different from the carrier decomposition in
Eqs.~\eqref{eq:jCO2_gas_def}--\eqref{eq:jCO2_na_def}, which separates the total
CO$_2$ current into gas-phase, aqueous-phase, and nonadvective contributions.
Here, by contrast, \(\mathbf j_{\mathrm{tf}}\), \(\mathbf j_c\), and
\(\mathbf j_{\mathrm{grav}}\) are not additional phase currents. They are the
total-flow, capillary, and gravitational parts of
\(\mathbf j_{\mathrm{CO_2}}^g\).

The first term,
\(\mathbf j_{\mathrm{tf}}=n_g f_g\mathbf v_t\), carries the part of the mobile
gas-phase CO$_2$ current associated with the total Darcy flow. At the reduced
plume scale this contribution is represented by an effective pressure-field
redistribution closure after vertical averaging and projection onto \(u\). The
second term is the capillary relative-motion contribution. The third term is
the gravitational relative-motion contribution. No vertical average and no
projection onto \(u\) have been applied at this stage.

We now apply vertical reduction to the conservative gas-phase CO$_2$ balance, Eq.~\eqref{eq:nCO2_gas_balance_general}. Let \(\nabla_h\) denote the horizontal gradient and define the thickness average by \(\langle f\rangle=H^{-1}\int_0^H f\,dz\). Assuming no gas-phase CO$_2$ current through the top and bottom boundaries, thickness averaging gives
\begin{equation}
\frac{\partial n_{\mathrm{CO_2}}^{g,2D}}{\partial t}+\nabla_h\!\cdot\mathbf j^{2D}=\gamma^{2D},
\label{eq:gCO2_molar_balance_2D_general}
\end{equation}
where \(n_{\mathrm{CO_2}}^{g,2D}=\langle n_{\mathrm{CO_2}}^{g}\rangle\), \(\mathbf j^{2D}=\langle\mathbf j_{\mathrm{CO_2},h}^{g}\rangle\), and \(\gamma^{2D}=\langle\gamma_g\rangle\).

Vertical segregation replaces the detailed saturation profile by a CO$_2$-rich layer of thickness \(h(\mathbf x,t)\) beneath the caprock. We define
\begin{equation}
u(\mathbf x,t)=\frac{1}{H n_\ast}\int_0^H n_{\mathrm{CO_2}}^{g}(\mathbf x,z,t)\,dz=\frac{n_{\mathrm{CO_2}}^{g,2D}}{n_\ast},
\label{eq:theta_plume_variable_def}
\end{equation}
with \(0\le u\le1\). Here \(n_\ast\) is the reference gas-phase CO$_2$ molar concentration per bulk volume associated with a fully developed connected gas layer. Thus \(u\) is a bounded, vertically averaged measure of mobile free-phase CO$_2$, normalized by its maximum possible value in the aquifer column.

If \(S_{br}\) is the residual brine saturation inside the CO$_2$-rich layer, the vertically segregated approximation gives \(u=h/H\) and \(\overline S_g^{\,\mathrm{VE}}=(1-S_{br})u\). Here \(\overline S_g^{\,\mathrm{VE}}\) is the vertically averaged gas saturation, not the local gas saturation inside the CO$_2$-rich layer. For compactness, define \(\phi_b=\phi(1-S_{br})\). For a representative plume-scale gas molar concentration \(n_g^\ast\), \(n_\ast=\phi_b n_g^\ast\). Using the pure-gas relation above, the vertically reduced gas-phase CO$_2$ molar concentration is then
\begin{equation}
n_{\mathrm{CO_2}}^{g,2D}=n_\ast u .
\label{eq:nCO2_gas_2D_effective}
\end{equation}

We now vertically reduce the three mechanism-level contributions in
Eq.~\eqref{eq:jCO2_gas_mechanism_split} and project them onto the scalar plume
variable \(u\). The total-flow contribution is closed at plume scale by
\begin{equation}
\mathbf j_{\mathrm{tf}}^{2D}
\simeq
-n_\ast D_p(u)\nabla_h u .
\label{eq:jtf_2D_projected}
\end{equation}
Here \(D_p(u)\) is an effective pressure-field redistribution coefficient. It
summarizes the horizontal spreading produced by the total Darcy flow after
vertical averaging. It is therefore a plume-scale closure, not a separate
phase-current decomposition.

The capillary contribution is obtained from
\[
\mathbf j_c^{2D}
=
-D_c^{(p)}\nabla_h p_c^{\mathrm{VE}},
\qquad
D_c^{(p)}
=
n_g k\frac{\lambda_g\lambda_w}{\Lambda_t}.
\]
Using
\(\nabla_h p_c^{\mathrm{VE}}
=
(d p_c^{\mathrm{VE}}/du)\nabla_h u\), we write
\begin{equation}
\mathbf j_c^{2D}
=
-n_\ast D_c(u)\nabla_h u,
\qquad
D_c(u)
=
\frac{k n_g}{n_\ast}
\frac{\lambda_g\lambda_w}{\Lambda_t}
\frac{d p_c^{\mathrm{VE}}}{du}.
\label{eq:Dc_projected}
\end{equation}

The gravitational body-force term produces horizontal spreading only after
vertical segregation. A lateral variation of plume thickness gives the
hydrostatic driving gradient
\(\Delta\rho\,g\nabla_h h\). Since \(h=Hu\), this gives
\begin{equation}
\mathbf j_{\mathrm{grav}}^{2D}
=
-n_\ast D_{\mathrm{grav}}(u)\nabla_h u,
\quad
D_{\mathrm{grav}}(u)
=
\frac{k n_g H\Delta\rho\,g}{n_\ast}
\frac{\lambda_g\lambda_w}{\Lambda_t}.
\label{eq:Dgrav_projected}
\end{equation}

The projected horizontal current is therefore
\begin{equation}
\mathbf j^{2D}
=
\mathbf j_{\mathrm{tf}}^{2D}
+
\mathbf j_c^{2D}
+
\mathbf j_{\mathrm{grav}}^{2D}
=
-n_\ast D_u(u)\nabla_h u ,
\label{eq:j_effective_final}
\end{equation}
with
\begin{equation}
D_u(u)
=
D_p(u)
+
D_c(u)
+
D_{\mathrm{grav}}(u).
\label{eq:Du_decomposition_projected}
\end{equation}
Thus \(D_u\) is a plume-scale transport coefficient obtained after vertical
averaging and projection of the mobile gas-phase CO$_2$ current onto \(u\).
The terms \(D_p\), \(D_c\), and \(D_{\mathrm{grav}}\) represent, respectively,
the effective total-flow/pressure-field redistribution, capillary
redistribution, and gravity-current spreading.

Substitution of Eqs.~\eqref{eq:nCO2_gas_2D_effective} and \eqref{eq:j_effective_final} into Eq.~\eqref{eq:gCO2_molar_balance_2D_general} gives, for constant \(n_\ast\),
\begin{equation}
\frac{\partial u}{\partial t}=\nabla_h\!\cdot\left(D_u\nabla_h u\right)+\frac{\gamma^{2D}}{n_\ast}.
\label{eq:theta_effective_source}
\end{equation}

Here \(u(\mathbf x,t)\) is the dimensionless plume scale variable measuring
the vertically reduced amount of mobile  gas phase CO$_2$. The term
\(\nabla_h\!\cdot(D_u\nabla_h u)\) describes lateral plume spreading. The
effective coefficient \(D_u\) contains the combined effects of pressure driven
flow, buoyancy driven gravity segregation, and capillary redistribution. The
term \(\gamma^{2D}\) is the vertically averaged net molar source of mobile
 gas phase CO$_2$. It includes CO$_2$ added by injection, minus the CO$_2$
removed from the mobile gas phase by dissolution into brine and by
mineralization or other rock associated trapping processes. The division by
\(n_\ast\) converts this net molar source into the corresponding rate of
change of the normalized variable \(u\).

\subsection{Effective nonlinear current and the GBL-derived nonhomogeneous porous-medium equation}
\label{sec:gbl_pme}

Equation~\eqref{eq:theta_effective_source} is the reduced plume equation
obtained after vertical averaging and projection of the GBL/Darcy transport
structure onto the bounded scalar variable \(u\). It already contains two
distinct contributions: the nonlinear lateral redistribution of mobile
 gas phase CO$_2$ through the effective current, and the nonhomogeneous
source/sink term that accounts for exchange of mobile CO$_2$ with injection,
dissolution, mineralization, and retention processes.

The effective coefficient \(D_u\) appearing in
Eq.~\eqref{eq:theta_effective_source} has already been obtained from the
projected current decomposition in Eq.~\eqref{eq:j_effective_final}. It is
not a microscopic material diffusivity. It is a plume-scale transport
coefficient obtained after vertical averaging and projection of the
Darcy/GBL current onto the scalar variable \(u\).
Its mechanism-level components are defined by
Eqs.~\eqref{eq:jtf_2D_projected}, \eqref{eq:Dc_projected}, and
\eqref{eq:Dgrav_projected}.

The nonhomogeneous term in Eq.~\eqref{eq:theta_effective_source} must be
carried through the reduced model. The quantity \(\gamma^{2D}\) is the
vertically averaged net molar source of mobile  gas phase CO$_2$. It includes
both injection and the processes that remove CO$_2$ from the mobile gas phase:
\begin{equation}
\gamma^{2D}
=
\gamma_{\mathrm{inj}}^{2D}
-
\gamma_{\mathrm{diss}}^{2D}
-
\gamma_{\mathrm{min}}^{2D}
-
\gamma_{\mathrm{ret}}^{2D}.
\label{eq:gamma_2D_decomposition}
\end{equation}
Here \(\gamma_{\mathrm{inj}}^{2D}\) is the injection contribution,
\(\gamma_{\mathrm{diss}}^{2D}\) is the loss of mobile CO$_2$ by dissolution
into the aqueous phase, \(\gamma_{\mathrm{min}}^{2D}\) denotes
mineralization, and \(\gamma_{\mathrm{ret}}^{2D}\) denotes residual trapping
or other rock associated retention mechanisms. The injection contribution is
typically localized near the well, whereas dissolution is strongest near the
gas--brine contact and along the plume edge, where mobile CO$_2$ remains in
contact with undersaturated brine. The factor \(n_\ast\) in
Eq.~\eqref{eq:theta_effective_source} converts this net molar source into the
corresponding rate of change of the dimensionless plume variable \(u\).
The pressure-field contribution is treated as an
effective plume-scale closure, while the capillary and gravitational terms are
obtained by projecting the corresponding relative-motion terms onto \(u\).

To obtain an analytically tractable reduced model, we approximate the
effective spreading coefficient by the state dependent power law
\begin{equation}
D_u(u)
\simeq
D_0 u^{1-q},
\qquad
D_0>0,
\qquad
0\le q\le 1 .
\label{eq:Du_powerlaw}
\end{equation}
Here \(D_0\) has units of length squared per time, since \(u\) is
dimensionless. The parameter \(q\) is the nonlinear transport index inherited
from the nonextensive, Tsallis-type nonlinear diffusion closure
\cite{TsallisBukman1996}. It controls how the projected
plume scale spreading coefficient changes with the reduced  gas phase
CO$_2$ inventory. The limit \(q=0\) gives \(D_u(u)\simeq D_0u\), corresponding
to the constant compressibility nonlinear spreading limit. The limit \(q=1\)
gives \(D_u(u)=D_0\), corresponding to the ideal gas or linear diffusion
limit \cite{Vazquez2007PME}. 
For \(q<1\), the spreading coefficient $D_u$ vanishes as \(u\to0\); in this restricted
PDE sense the nonlinear operator is degenerate at the plume edge.

Substitution of Eq.~\eqref{eq:Du_powerlaw} into
Eq.~\eqref{eq:theta_effective_source} gives
\begin{equation}
\frac{\partial u}{\partial t}
=
D_0
\nabla_h\!\cdot
\left(
u^{1-q}
\nabla_h u
\right)
+
\frac{\gamma^{2D}}{n_\ast}.
\label{eq:nonhomogeneous_PME_cartesian_u}
\end{equation}
This is the GBL-derived nonhomogeneous porous-medium equation in Cartesian
form. The nonlinear divergence term represents the projected lateral spreading
current, while the source/sink term remains explicitly present and carries the
net gain or loss of mobile  gas phase CO$_2$.
For \(0\le q<1\), Eq.~\eqref{eq:nonhomogeneous_PME_cartesian_u} 
belongs to the nonlinear finite front spreading class because
\(D_u(u)\sim u^{1-q}\to 0\) as \(u\to0\). The endpoint \(q=1\) gives the linear diffusion
limit, for which the compact support Barenblatt profile is no longer literal.
In all cases, \(\gamma^{2D}\) retains the nonconservative exchange of mobile
 gas phase CO$_2$ with injection, dissolution, mineralization, and retention
processes.

For the axisymmetric CO$_2$ plume footprint considered below,
Eq.~\eqref{eq:nonhomogeneous_PME_cartesian_u} reduces to
\begin{equation}
\frac{\partial u}{\partial t}
=
\frac{D_0}{r}
\frac{\partial}{\partial r}
\left(
r u^{1-q}
\frac{\partial u}{\partial r}
\right)
+
\frac{\gamma^{2D}(r,t)}{n_\ast}.
\label{eq:nonhomogeneous_PME_axisymmetric_u}
\end{equation}
This is the axisymmetric GBL-derived reduced porous-medium equation with a
nonhomogeneous source/sink term for the normalized mobile  gas phase CO$_2$
plume variable. It carries both the nonlinear lateral spreading current and
the net source/sink structure of the storage problem. The injection
contribution mainly supplies mobile CO$_2$ near the well, whereas dissolution,
mineralization, and retention remove mobile CO$_2$ from the  gas phase plume
as it migrates and remains in contact with brine and reactive rock.

The parameters \(D_0\) and \(q\) are effective parameters of the reduced
plume scale model. The coefficient \(D_0\) fixes the lateral spreading scale. Their identification does not determine a
unique equation of state, relative permeability law, capillary pressure
relation, or pressure field. Rather, the pair \((D_0,q)\) summarizes the
combined nonlinear spreading response obtained after projection of the
Darcy/GBL transport structure onto the scalar variable \(u\), while
\(\gamma^{2D}\) carries the net addition and removal of mobile  gas phase
CO$_2$ by injection, dissolution, mineralization, and retention. The
Barenblatt-type analysis developed below should therefore be understood as the
spreading response of the nonlinear transport operator, with the source/sink
term retained at the reduced equation level and specified separately according
to the injection protocol and trapping model.

\subsection{Mass-controlled Barenblatt reference profile for the forced plume}
\label{sec:barenblatt}

We now characterize the compact spreading structure associated with the
axisymmetric forced reduced equation
\eqref{eq:nonhomogeneous_PME_axisymmetric_u}. The source/sink term is part of
the reduced storage model. It represents the net gain or loss of mobile
 gas phase CO$_2$ by injection, dissolution, mineralization, and retention.
Consequently, the mobile plume content is not generally conserved.

We define the reduced mobile content as
\begin{equation}
M_u(t)
=
2\pi
\int_0^\infty
u(r,t)\,r\,dr .
\label{eq:Mu_axisymmetric_def}
\end{equation} The corresponding
physical mobile  gas phase CO$_2$ volume is
\begin{equation}
V_{\mathrm{CO_2}}^{\mathrm{mob}}(t)
=
\phi(1-S_{br})H\,M_u(t)
 .
\label{eq:Vco2_u_definition}
\end{equation}
This quantity is not the cumulative injected CO$_2$ volume. It is the volume
of CO$_2$ that remains in the mobile  gas phase plume after injection,
dissolution, mineralization, residual trapping, and retention have acted.
Integrating Eq.~\eqref{eq:nonhomogeneous_PME_axisymmetric_u} over the
axisymmetric plume footprint gives
\begin{equation}
\frac{dM_u}{dt}
=
\frac{2\pi}{n_\ast}
\int_0^\infty
\gamma^{2D}(r,t)\,r\,dr ,
\label{eq:Mu_axisymmetric_balance}
\end{equation}
provided that the nonlinear current gives no net contribution at the external
boundary. Thus, injection increases \(M_u\), whereas dissolution,
mineralization, and retention decrease it.

Instead of prescribing the pointwise structure of
\(\gamma^{2D}(r,t)\), we represent its net effect through the mobile amount
law
\begin{equation}
M_u(t)
=
M_{u0}
+
\mathcal A t^\alpha ,
\qquad
\mathcal A\ge0,
\qquad
\alpha\ge0 .
\label{eq:Mu_powerlaw_axisymmetric}
\end{equation}
Equivalently, using Eq.~\eqref{eq:Vco2_u_definition}, the physical mobile
 gas phase CO$_2$ volume satisfies
\begin{equation}
V_{\mathrm{CO_2}}^{\mathrm{mob}}(t)=V_{\mathrm{CO_2},0}^{\mathrm{mob}}+\mathcal Q_\alpha t^\alpha .
\label{eq:Vco2_mob_powerlaw}
\end{equation}
Here \(V_{\mathrm{CO_2},0}^{\mathrm{mob}}=\phi(1-S_{br})H\,M_{u0}\) is the initial mobile  gas phase CO$_2$ volume, and \(\mathcal Q_\alpha=\phi(1-S_{br})H\,\mathcal A\) is the corresponding physical quantity growth coefficient.
For \(\alpha=1\), \(\mathcal Q_\alpha\) has units of volume per time and
represents the net growth rate of the mobile  gas phase CO$_2$ plume. This
rate need not be identical to the imposed injection rate if dissolution,
mineralization, residual trapping, or leakage remove CO$_2$ from the mobile
gas phase.

Here \(\alpha\) is the growth exponent of the  gas phase CO$_2$ amount, not
directly the imposed well rate exponent. The case \(\alpha=0\) corresponds to
an approximately conserved  gas phase CO$_2$ inventory, as in a shut-in regime
with negligible post-injection losses. The case \(\alpha=1\) corresponds to
linear inventory growth, which is obtained for constant imposed CO$_2$
injection when dissolution, mineralization, residual trapping, and leakage do
not alter the leading-order time dependence of the  gas phase inventory.
Sublinear growth, \(0<\alpha<1\), represents a  gas phase inventory that grows
more slowly than the injected amount because sinks remove part of the CO$_2$.
Superlinear growth, \(\alpha>1\), represents accelerated growth of the
 gas phase content.

The approximation used below is mass-controlled. The source/sink term fixes
\(M_u(t)\), while the nonlinear GBL-derived spreading operator fixes the
compact radial shape. We therefore use the Barenblatt profile of the
homogeneous nonlinear operator with its mass replaced by the time-dependent
inventory in Eq.~\eqref{eq:Mu_powerlaw_axisymmetric}. This is not an exact
pointwise solution of the forced equation
\eqref{eq:nonhomogeneous_PME_axisymmetric_u}; it neglects the detailed
spatial distribution of \(\gamma^{2D}(r,t)\). It is intended as a
plume radius scaling approximation and as a reference profile for the
nonlinear spreading structure.

Due to the homogeneous operator of the Eq.~\eqref{eq:nonhomogeneous_PME_cartesian_u} the axisymmetric Barenblatt-type profile can be written as
\begin{equation}
u(r,t)
=
A_B(t)
\left[
1-\frac{r^2}{R^2(t)}
\right]_+^{1/(1-q)} ,
\label{eq:axisymmetric_barenblatt_mass_profile}
\end{equation}
where \(A_B(t)\) is the central amplitude and \(R(t)\) is the compact support
radius. The positive part operator sets \(u=0\) for \(r\ge R(t)\).

Define
\begin{equation}
\kappa
=
\frac{1-q}{4D_0(2-q)} .
\label{eq:kappa_axisymmetric}
\end{equation}
The mass-controlled radius and amplitude are then
\begin{equation}
R(t)
=
\kappa^{-1/2}
t^{1/[2(2-q)]}
\left[
\frac{M_u(t)}{4\pi D_0}
\right]^{
(1-q)/[2(2-q)]
},
\label{eq:R_mass_controlled_axisymmetric}
\end{equation}
and
\begin{equation}
A_B(t)
=
\left[
\frac{M_u(t)}{4\pi D_0t}
\right]^{1/(2-q)} .
\label{eq:AB_mass_controlled}
\end{equation}
With these definitions,
\begin{equation}
2\pi
\int_0^{R(t)}
u(r,t)\,r\,dr
=
M_u(t),
\label{eq:barenblatt_mass_normalization_axisymmetric}
\end{equation}
so the reference profile is exactly consistent with the prescribed mobile
quantity.

Substitution of Eq.~\eqref{eq:Mu_powerlaw_axisymmetric} into
Eq.~\eqref{eq:R_mass_controlled_axisymmetric} gives
\begin{equation}
R(t)
=
\kappa^{-1/2}
t^{1/[2(2-q)]}
\left[
\frac{M_{u0}+\mathcal A t^\alpha}{4\pi D_0}
\right]^{
(1-q)/[2(2-q)]
}.
\label{eq:R_powerlaw_inventory_axisymmetric}
\end{equation}
Similarly, the amplitude becomes
\begin{equation}
A_B(t)
=
\left[
\frac{M_{u0}+\mathcal A t^\alpha}{4\pi D_0t}
\right]^{1/(2-q)} .
\label{eq:AB_powerlaw_inventory}
\end{equation}

Equation~\eqref{eq:R_powerlaw_inventory_axisymmetric} contains the radius
scaling for a general power-law mobile inventory.
If the plume starts from negligible mobile inventory,
\begin{equation}
M_{u0}=0,
\label{eq:zero_initial_inventory}
\end{equation}
then Eq.~\eqref{eq:Mu_powerlaw_axisymmetric} gives
\(M_u(t)=\mathcal A t^\alpha\). More generally, the same scaling is obtained
asymptotically whenever
\begin{equation}
\mathcal A t^\alpha\gg M_{u0}.
\label{eq:inventory_growth_regime}
\end{equation}
In both cases, Eq.~\eqref{eq:R_powerlaw_inventory_axisymmetric} gives
\begin{equation}
R(t)
\sim
t^{\beta_\alpha},
\qquad
\beta_\alpha
=
\frac{1+\alpha(1-q)}{2(2-q)} .
\label{eq:beta_alpha_axisymmetric}
\end{equation}
Thus the observed radius exponent depends both on the nonlinear transport
parameter \(q\) and on the mobile content exponent \(\alpha\). In the same
regime, Eq.~\eqref{eq:AB_powerlaw_inventory} gives
\begin{equation}
A_B(t)
\sim
t^{(\alpha-1)/(2-q)} .
\label{eq:AB_scaling_alpha}
\end{equation}

The opposite limiting case is an initially present mobile plume with no
further net mobile addition,
\begin{equation}
M_{u0}\neq0,
\qquad
\mathcal A=0 .
\label{eq:shutin_inventory_limit}
\end{equation}
This corresponds to the shut-in or approximately conserved quantity regime.
Then \(M_u(t)=M_{u0}\), and Eq.~\eqref{eq:R_powerlaw_inventory_axisymmetric}
reduces to
\begin{equation}
R(t)
\sim
t^{1/[2(2-q)]}.
\label{eq:conserved_inventory_radius_scaling}
\end{equation}
For \(0\le q<1\), this exponent lies between \(1/4\) and \(1/2\). The endpoint
\(q=1\) corresponds to the linear diffusion limit and should be interpreted
through a defined plume threshold radius rather than a compact support
Barenblatt front.

For constant imposed CO$_2$ injection, the cumulative injected amount is
linear in time. If dissolution, mineralization, residual trapping, and other
loss mechanisms do not change this leading order inventory scaling, then the
 gas phase CO$_2$ content also grows linearly, corresponding to
\(\alpha=1\). Equation~\eqref{eq:beta_alpha_axisymmetric} then gives
\[
R(t)\sim t^{1/2},
\]
independently of \(q\). Therefore a square-root plume radius law does not by
itself imply linear diffusion; it can also arise from nonlinear
Barenblatt-type spreading with linearly growing  gas phase CO$_2$ inventory
\cite{Barenblatt1952,Pattle1959,Vazquez2007PME}. More generally,
sublinear quantity growth, \(0<\alpha<1\), gives
\(\beta_\alpha<1/2\), whereas superlinear amount growth, \(\alpha>1\),
gives \(\beta_\alpha>1/2\).

In the constant compressibility limit \(q=0\),
Eq.~\eqref{eq:beta_alpha_axisymmetric} reduces to
\[
\beta_\alpha=\frac{1+\alpha}{4}.
\]
Thus an approximately constant mobile inventory gives \(R(t)\sim t^{1/4}\),
whereas constant net mobile injection gives \(R(t)\sim t^{1/2}\). In the
formal limit \(q\to1\), the equation approaches linear diffusion. The
compact support Barenblatt profile is then no longer literal, and the plume
radius must be defined by a concentration or saturation threshold; the
characteristic diffusive length nevertheless scales as \(t^{1/2}\).

This mass-controlled construction separates the role of transport from the
role of forcing. The nonlinear GBL-derived operator determines the compact
radial shape and the \(q\)-dependent spreading response, while the
nonhomogeneous source/sink term determines the mobile inventory \(M_u(t)\).
When \(M_u(t)\sim t^\alpha\), the corresponding plume-radius scaling is
\[
R(t)\sim
t^{[1+\alpha(1-q)]/[2(2-q)]}.
\]
This relation contains both the conserved quantity Barenblatt limit and the
constant injection limit relevant to field scale plume interpretation.

\subsection{Capped Barenblatt-type plume with a finite full-thickness core}
\label{sec:capped_barenblatt}

The Barenblatt-type reference profile in Sec.~\ref{sec:barenblatt} describes
the compact spreading structure generated by the nonlinear GBL-derived
transport operator. The physical plume variable, however, is bounded by
\(0\le u\le 1\), because \(u=h/H\) in the vertically segregated limit. If the
unconstrained Barenblatt-type profile would exceed unity near the injection
region, the reduced plume develops a full-thickness core. We therefore use a
capped profile with two moving radii: the core radius \(a(t)\), where
\(u=1\), and the outer mobile-plume edge \(R(t)\), where \(u=0\).

The capped Barenblatt-type profile is
\begin{equation}
u(r,t)
=
\begin{cases}
1, & 0\le r\le a(t), \\[4pt]
\left[
\dfrac{R(t)^2-r^2}{R(t)^2-a(t)^2}
\right]^{1/(1-q)},
& a(t)<r<R(t), \\[10pt]
0, & r\ge R(t).
\end{cases}
\label{eq:capped_barenblatt_profile}
\end{equation}
The exponent \(1/(1-q)\) is inherited from the Barenblatt-type tail of the
GBL-derived nonlinear transport operator. The new ingredient is the upper
bound \(u\le1\), which creates the ful-thickness core.

It is useful to work with the reduced mobile area
\begin{widetext}
    \begin{equation}
A_u(t)
\equiv
\frac{V_{\mathrm{CO_2}}^{\mathrm{mob}}(t)}
{\pi\phi(1-S_{br})H}
=
\frac{M_u(t)}{\pi}
=
a(t)^2
+
\frac{1-q}{2-q}
\left[
R(t)^2-a(t)^2
\right].
\label{eq:Au_capped}
\end{equation}
\end{widetext}

The first term is the ful-thickness core contribution, while the second term
is the Barenblatt-type transition-region contribution. The quantity
\(V_{\mathrm{CO_2}}^{\mathrm{mob}}\) is not the cumulative injected CO$_2$
volume; it is the amount that remains in the mobile gas phase after injection,
dissolution, mineralization, residual trapping, and retention have acted.

The time dependence of \(A_u(t)\) is inherited from the mobile inventory
\(M_u(t)\). If the integrated source/sink history is represented by
Eq.~\eqref{eq:Mu_powerlaw_axisymmetric}, then
\begin{equation}
A_u(t)
=
A_{u0}
+
\mathcal B t^\alpha,
\qquad
A_{u0}=\frac{M_{u0}}{\pi},
\qquad
\mathcal B=\frac{\mathcal A}{\pi}.
\label{eq:Au_powerlaw_inventory}
\end{equation}
Thus the source/sink term is not removed from the capped construction; it
enters through the prescribed or modeled history of \(A_u(t)\).

The capped profile still requires one relation between \(a(t)\) and \(R(t)\).
We take this relation from the radial scale of the nonlinear Barenblatt-type
tail:
\begin{equation}
R(t)^2-a(t)^2
=
\frac{4D_0(2-q)}{1-q}\,t .
\label{eq:tail_width_relation_forced}
\end{equation}
Combining this relation with Eq.~\eqref{eq:Au_capped} gives
\begin{align}
a(t)^2
&=
A_u(t)-4D_0t,
\\
R(t)^2
&=
A_u(t)+\frac{4D_0}{1-q}\,t . 
\label{eq:a2_R2_general_capped}
\end{align}
Using Eq.~\eqref{eq:Au_powerlaw_inventory}, this becomes
\begin{align}
a(t)^2
&=
A_{u0}
+
\mathcal B t^\alpha
-
4D_0t,
\\
R(t)^2
&=
A_{u0}
+
\mathcal B t^\alpha
+
\frac{4D_0}{1-q}t .
\label{eq:a2_R2_powerlaw_capped}
\end{align}
The capped regime exists while \(a(t)^2>0\). When \(a(t)\) reaches zero, the
ful-thickness core disappears and the plume crosses over to a single-front
Barenblatt-type profile. This crossover is geometric; it does not imply that
the source/sink term has vanished.

\section{Results}
\label{sec:results}

\begin{figure*}[t]
  \centering
  \includegraphics[width=0.7\linewidth,trim=0cm 0cm 0cm 0cm,clip]{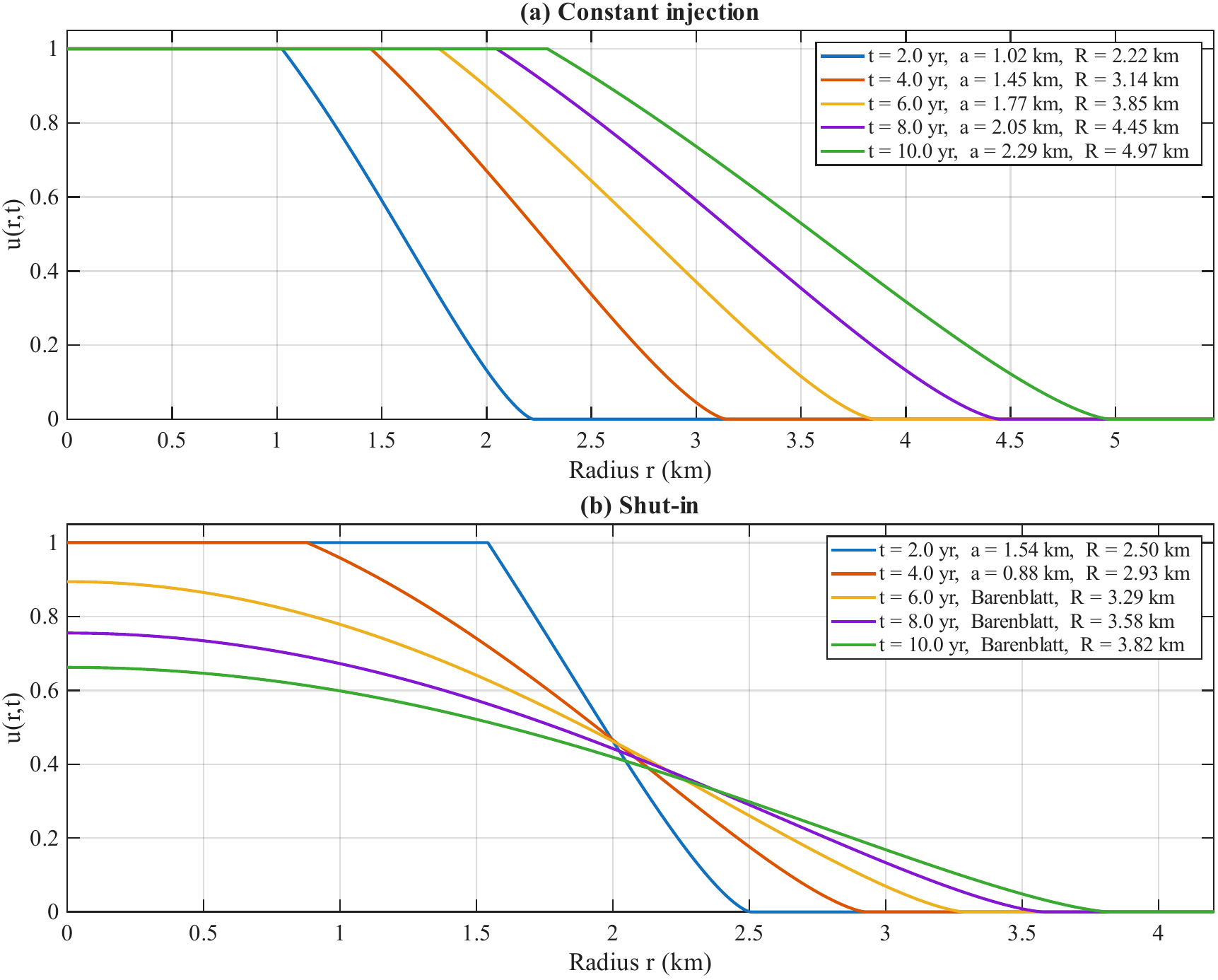}
  \caption{Bounded plume profiles \(u(r,t)\) for the finite core model, plotted as a function of radius in km. 
  (a) Constant net mobile injection: the plume develops an inner saturated region, \(u=1\) for \(r\le a(t)\), followed by a Barenblatt-type outer tail that vanishes at the plume edge \(r=R(t)\). As time increases, both the core radius \(a(t)\) and the plume edge \(R(t)\) move outward. 
  (b) Shut-in: after injection stops, the inner saturated core shrinks with time, while the outer plume edge can continue to advance by nonlinear redistribution of the remaining mobile  gas phase CO$_2$. Once \(a(t)\) reaches zero, the finite core disappears and the profile crosses over to a tail-only Barenblatt-type regime.}
  \label{fig:core_profiles_ab}
\end{figure*}

This section presents the main results of the study, combining analytical model predictions with field scale observations. First, we examine the evolution of finite core plume profiles derived from the composite solution, highlighting the distinct behaviors that arise under constant injection and after injection is paused. In particular, we show how the presence or disappearance of a ful-thickness inner core controls the transition between injection driven spreading and redistribution governed by nonlinear diffusion. Second, we compare these theoretical predictions with field evidence by reconstructing plume radius evolution from published time lapse monitoring data. The resulting scaling exponents are shown to be broadly consistent with the similarity solutions of the porous medium equation, providing support for interpreting field scale plume growth in terms of effective nonlinear diffusion dynamics.

\subsection{Evolution of finite core plume profiles under constant injection and after shut-in}
\label{subsec:core_profiles_ab}

Figure~\ref{fig:core_profiles_ab} illustrates the time evolution of the bounded plume variable \(u(r,t)\) for two idealized regimes. The figure is not a site specific history match; it is a numerical illustration of the capped analytical construction. In both panels, the profile contains an inner saturated core, \(u=1\) for \(0\le r\le a(t)\), joined continuously to a Barenblatt-type tail for \(a(t)<r<R(t)\), with \(u=0\) for \(r\ge R(t)\). Under vertical segregation, the plateau \(u=1\) corresponds to a gas  layer occupying the full available aquifer thickness, while the outer branch represents the partially filled transition region between the core and the mobile plume edge.

The parameters used to generate Fig.~\ref{fig:core_profiles_ab} are listed in Table~\ref{tab:core_profile_parameters}. All calculations were performed in metres, and the radial coordinate was converted to kilometres only for plotting.

\begin{table}[t]
\centering
\caption{Parameters used to generate the finite core profiles in Fig.~\ref{fig:core_profiles_ab}.}
\label{tab:core_profile_parameters}
\begin{tabular}{lll}
\hline
Quantity & Value & Description \\
\hline
\(H\) & \(10~\mathrm{m}\) & Aquifer thickness \\
\(\phi\) & \(0.30\) & Porosity \\
\(S_{br}\) & \(0.20\) & Residual brine saturation \\
\(q\) & \(0.30\) & Nonlinear transport index \\
\(D_0\) & \(2.0\times10^{4}~\mathrm{m^2\,yr^{-1}}\) & Spreading coefficient \\
\(Q_{\mathrm{mob}}\) & \(1.0\times10^{7}~\mathrm{m^3\,yr^{-1}}\) & Net mobile injection rate \\
\(V_0\) & \(3.0\times10^{7}~\mathrm{m^3}\) & Mobile volume at shut-in \\
\(t\) & \(2,4,6,8,10~\mathrm{yr}\) & Times shown in each panel \\
\hline
\end{tabular}
\end{table}

Panel (a) shows constant net mobile injection. In this case the mobile  gas phase CO$_2$ content increases linearly with time, \(V_{\mathrm{CO_2}}^{\mathrm{mob}}=Q_{\mathrm{mob}}t\). For the parameters in Table~\ref{tab:core_profile_parameters}, the supply is sufficient to maintain a finite saturated core, so both \(a(t)\) and \(R(t)\) increase with time. The radius \(a(t)\) marks the outer boundary of the region where \(u=1\), while \(R(t)\) marks the compact support edge where the mobile  gas phase plume vanishes. In this injection controlled capped regime, the outer radius follows the square-root scaling \(R(t)\propto t^{1/2}\).

Panel (b) shows the shut-in case. The initial mobile volume is fixed at \(V_0\), so no additional mobile CO$_2$ is supplied to sustain the saturated core. The core radius \(a(t)\) therefore decreases with time, while the outer radius \(R(t)\) can continue to advance as the existing mobile quantity redistributes laterally. When \(a(t)\) reaches zero, the condition \(u=1\) is no longer satisfied over a finite radial interval, and the capped profile crosses over to a single front Barenblatt-type profile. In the approximately conservative post core limit, the outer radius follows \(R(t)\propto t^{1/[2(2-q)]}\).

The figure therefore separates two geometric regimes of the bounded plume model. Constant net mobile injection can maintain a ful-thickness core and produce injection controlled square-root growth. After shut-in, the ful-thickness core is transient; once it disappears, the plume evolves as a tail only compact profile governed by nonlinear redistribution of the remaining mobile  gas phase CO$_2$ inventory.

\subsection{Application to time-lapse seismic plume data}
\label{sec:data_application}

This section develops an image based comparison of CO$_2$ plume evolution across three benchmark storage sites: Aquistore, Sleipner, and Weyburn--Midale. Aquistore provides a compact and well monitored saline aquifer plume, making it useful for testing apparent radial growth from repeated seismic images. Sleipner is the reference industrial scale offshore storage case, where injected CO$_2$ is distributed across several intra-reservoir layers; it is therefore especially valuable for studying vertical migration, layer-by-layer spreading, and delayed plume initiation. Weyburn--Midale extends the analysis to CO$_2$ storage coupled with enhanced oil recovery, where multiple injection and production wells generate spatially distributed seismic anomalies. For each site, published plume images are converted into detected footprint areas and area equivalent radii, with uncertainty estimated from image resolution and boundary extraction.

The purpose of this comparison is not to perform a full history match, but to test which reduced spreading regimes are consistent with the observed plume footprint evolution. We use the area equivalent radius
\begin{equation}
R_{eq}(t)
=
\left(
\frac{A(t)}{\pi}
\right)^{1/2},
\label{eq:area_equivalent_radius}
\end{equation}
where \(A(t)\) is the mapped plume area or an equivalent areal measure. For each plume mask, the area is computed as \(A=N_p\Delta x^2\), where \(N_p\) is the number of detected plume pixels and \(\Delta x\) is the calibrated pixel size. A one-pixel uncertainty is assigned along the detected boundary, giving \(\delta A=N_{\partial p}\Delta x^2\). Propagation through Eq.~\eqref{eq:area_equivalent_radius} gives
\begin{equation}
\delta R_{eq}
=
\frac{\delta A}{2\pi R_{eq}}.
\label{eq:area_equivalent_radius_uncertainty}
\end{equation}
These uncertainties describe pixel-scale boundary placement and thresholding; they do not represent the full seismic interpretation uncertainty.

For each site, the extracted radius sequence is fitted with
\begin{equation}
R_{eq}(t)
=
R_0(t-t_0)^\beta ,
\label{eq:radius_power_law_fit}
\end{equation}
where \(R_0\) sets the radial scale, \(t_0\) is an apparent onset time for the mapped seismic response, and \(\beta\) is the apparent footprint growth exponent. The fitted time \(t_0\) should not be interpreted as the physical start of CO$_2$ injection. It accounts for the delay between injection, detectable saturation change, survey timing, and the resolution of the published seismic maps. The exponent \(\beta\) is therefore a descriptor of the mapped seismic footprint, not a direct inversion of the nonlinear exponent \(q\) in the reduced plume equation.

We recall that the mass-controlled scaling in Eq.~\eqref{eq:Mu_powerlaw_axisymmetric} writes the reduced mobile free-phase CO$_2$ inventory as
\[
M_u(t)=M_{u0}+\mathcal A t^\alpha .
\]
The exponent \(\alpha\) therefore measures the time dependence of the mobile inventory: \(\alpha=0\) corresponds to an approximately conserved inventory, while \(\alpha=1\) corresponds to linear growth, as expected for constant net mobile injection. The associated plume-radius exponent is given by Eq.~\eqref{eq:beta_alpha_axisymmetric},
\[
R(t)\sim t^{\beta_\alpha}, \qquad \beta_\alpha=\frac{1+\alpha(1-q)}{2(2-q)} .
\]
Thus the fitted footprint exponent \(\beta\) cannot be interpreted as a direct measurement of \(q\) unless the inventory exponent \(\alpha\) is also known. In particular, \(\alpha=0\) recovers the source-free Barenblatt range \(1/4<\beta<1/2\) for \(0<q<1\), whereas \(\alpha=1\) gives the square-root law \(\beta=1/2\), independently of \(q\).

\subsubsection{Weyburn--Midale field}


The Weyburn--Midale field is located in southeastern Saskatchewan, western Canada, close to the Canada--United States border. It is a benchmark onshore CO$_2$ monitoring site because CO$_2$ was injected into the oil reservoir for enhanced oil recovery (EOR) while oil was produced from nearby production wells. Field scale CO$_2$ injection began in the Weyburn Field in 2000, and the IEA Weyburn--Midale project used this commercial operation to test monitoring methods for tracking subsurface CO$_2$ migration and storage during EOR \cite{White2009Weyburn}. The image analysis in this study uses the Midale Marly time-lapse amplitude difference maps reported by White \cite[Fig.~5]{White2009Weyburn}. These maps cover a survey area containing several dual-leg horizontal wells, including both CO$_2$ injection wells and oil-production wells. The mapped anomaly is therefore treated as a distributed seismic plume footprint, not as a plume generated by a single injection point.

The plume footprint was extracted directly from the published seismic figure using colour based segmentation. The original figure was first calibrated using the scale bar. The 2002, 2004, and 2007 panels were then cropped, and a polygonal survey boundary was manually drawn around each mapped area to exclude labels, axes, and surrounding white space. Within each polygon, the RGB image was converted to HSV colour space, and plume pixels were selected from the warm hue range corresponding to the yellow--orange--red seismic anomaly. Additional saturation and brightness thresholds were used to reject grey, black, and weakly coloured pixels, including well symbols and background annotations. The resulting binary mask was cleaned by removing small isolated components, applying a mild morphological closing, and filling internal holes. All remaining connected components were retained because the Weyburn--Midale seismic response reflects a multiwell injection--production pattern rather than a single connected plume. 

 For Weyburn--Midale, \(\beta\) should be interpreted as a field-scale descriptor of the mapped anomaly rather than as the exponent of a single axisymmetric plume, because the seismic footprint is influenced by multiple CO$_2$ injection wells, oil production wells, and operational heterogeneity.

\subsubsection{Aquistore dedicated CO$_2$ storage project}
\label{subsec:aquistore_image_plumes}

The Aquistore site is a dedicated geological CO$_2$ storage project located near Estevan in southeastern Saskatchewan, western Canada, close to the Boundary Dam power station. Unlike Weyburn--Midale, Aquistore is not an EOR project: captured CO$_2$ is injected into a deep saline reservoir for storage rather than to produce oil. Injection began in April 2015, and repeated time lapse seismic surveys have been used to monitor the development of the subsurface plume \citep{RoachWhite2018}. The upper Deadwood panels analyzed here correspond to cumulative injected amounts of 36, 102, 141, 272, and 566 kt, with the latest survey acquired in November 2023 \cite[Fig.~1]{WhiteMovahedzadeh2025}. These values identify the injection state associated with each seismic image; they are not used here as a continuous injection rate history.

The image analysis was performed directly on the published upper Deadwood time lapse amplitude difference panels of \cite[Fig.~1]{WhiteMovahedzadeh2025}. The extracted object is therefore a figure based seismic footprint, not an independent inversion of CO$_2$ saturation. The image scale was calibrated from the 500 m reference circle in the source figure, giving approximately 4.31 m per pixel. Annotations, labels, and parts of the reference circle overlapping the panels were removed before segmentation by local neighbourhood interpolation from the surrounding image colours. The panel labelled 150 kt in the figure was retained with that visual label, but the quantitative injected amount used in the analysis was 141 kt, following the value reported in the text of \cite[Fig.~1]{WhiteMovahedzadeh2025}.

The plume footprint was segmented from the cleaned panels using a fixed blue dominance rule applied to the RGB channels. Pixels were classified as plume pixels when the blue channel exceeded a fixed fraction of its panel wise maximum and the red and green channels remained below prescribed fractions of the blue value. The resulting masks were cleaned by removing small isolated components, applying a mild morphological closing, filling holes, and retaining the largest connected plume patch. For the 272 kt panel, the large pale region inside the plume was added by selecting the largest white or weakly saturated cluster located inside or adjacent to the blue mask. This prevents the central anomaly from being excluded solely because it appears pale rather than blue in the published rendering. 

\subsubsection{Topmost Sleipner plume layer}
\label{subsec:sleipner_top_layer}

The Sleipner storage project is one of the most heavily documented industrial scale $\mathrm{CO}_2$ storage sites. It is located offshore Norway in the North Sea, where $\mathrm{CO}_2$ separated from produced natural gas has been injected since 1996 into the brine saturated Utsira Sand, a shallow saline aquifer rather than an oil reservoir. The injected $\mathrm{CO}_2$ rises buoyantly through the permeable sand and is redistributed beneath thin intra-reservoir mudstone layers, producing a vertically tiered plume structure that has been repeatedly imaged by time lapse seismic surveys. Here we focus only on the topmost mapped layer in the image compilation of Boait et al.~\citep[Fig.~4]{Boait2012Sleipner}, which provides a sequence of plume footprints for the years 1999, 2001, 2002, 2004, 2006, and 2008.

The image analysis was performed directly on the published panels of Boait et al.~\citep[Fig.~4]{Boait2012Sleipner}. The topmost layer panels were cropped from the original figure and segmented using a colour based threshold in HSV space. Pixels with sufficiently high colour saturation and brightness were retained as plume pixels, small isolated components were removed, and the mask was mildly closed and hole filled to obtain a connected image derived footprint.
Only the plume footprint identified through the image segmentation is considered in the analysis, while the remaining portions of the seismic interpretation are treated as background. Since the seismic interpretation panels were already denoised prior to this work, parts of the $\mathrm{CO}_2$ plume below the effective image or seismic detection threshold, including regions thinner than approximately $1~\mathrm{m}$, may have been removed before the segmentation was carried out.


\subsection{Comparison with the reduced-model regimes}
\label{sec:comparison_reduced_regimes}

\begin{table}[b]
\centering
\caption{
Power-law fit parameters for \(R_{eq}=R_0(t-t_0)^\beta\). The fitted onset time \(t_0\) is constrained to lie after injection start and before the first mapped plume observation.
}
\label{tab:model_data_comparison}
\begin{tabular}{lccccc}
\hline
Site & Injection start & Fitted \(t_0\) & \(R_0\) (km) & \(\beta\) & \(R^2_{\log}\) \\
\hline
Sleipner & 1996.00 & 1998.50 & 0.159 & 0.665 & 0.999 \\
Aquistore & 2015.25 & 2015.25 & 0.128 & 0.449 & 0.992 \\
Weyburn & 2000.00 & 2000.00 & 0.759 & 0.478 & 0.994 \\
\hline
\end{tabular}
\end{table}

\begin{figure}[t]
    \centering
    \includegraphics[width=\linewidth,trim={0.5cm 4.5cm 1cm 6cm},clip]{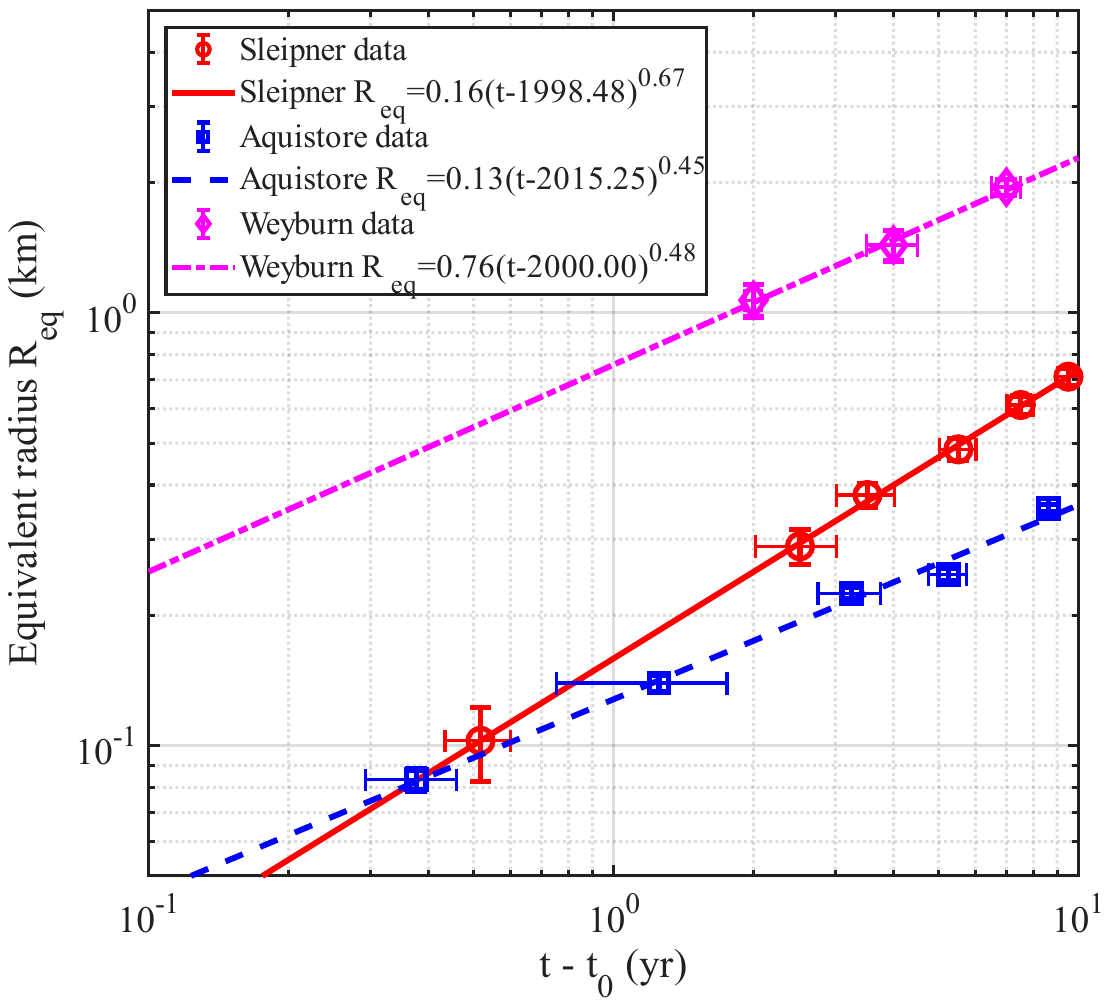}
    \caption{
    Equivalent radius evolution for the three monitored sites, plotted against the fitted time coordinate \(t-t_0\). 
    Symbols show image derived plume radii, and lines show the fitted laws \(R_{eq}=R_0(t-t_0)^\beta\). 
    Sleipner corresponds only to the topmost mapped layer. 
    Aquistore uses all five upper Deadwood panels. 
    Weyburn--Midale uses the plume areas extracted from the published Fig.~5 seismic maps. 
    Vertical error bars represent the propagated radius uncertainty from the image derived plume area uncertainty. 
    Horizontal error bars represent uncertainty in the survey timing.
    }
    \label{fig:all-sites-radius-growth}
\end{figure}

Figure~\ref{fig:all-sites-radius-growth} compares the image derived equivalent radius evolution for Aquistore, Weyburn--Midale, and the topmost Sleipner layer. The same fitting form,
\[
R_{eq}=R_0(t-t_0)^\beta ,
\]
is used for all three sites, and the fitted parameters are reported in Table~\ref{tab:model_data_comparison}. The text below therefore emphasizes the interpretation of the fitted exponent \(\beta\) and of the apparent onset time \(t_0\), rather than repeating the numerical values listed in the table.

The main observation is that Aquistore and Weyburn--Midale lie close to the square-root reference \(R_{eq}\propto (t-t_0)^{1/2}\). In the reduced model, this scaling is associated with injection controlled spreading of a mobile plume whose effective mobile mass continues to grow. Aquistore falls slightly below the square-root reference, consistent with a plume footprint whose areal growth is slower than linear over the monitoring interval. Weyburn--Midale is closer to the square-root trend, but its interpretation is different: the mapped anomaly reflects a multiwell injection--production pattern rather than a single axisymmetric plume. Thus its exponent should be read as an effective footprint growth exponent for the seismic anomaly, not as a direct measurement of a single plume spreading law.

The Sleipner topmost layer behaves differently. Its fitted exponent is larger than the square-root reference, and its fitted onset time is delayed relative to the start of injection. This delay is physically significant. It reflects the fact that the topmost layer did not respond immediately when injection began; CO$_2$ first migrated through the lower parts of the Utsira Sand and filled lower intra-reservoir traps before reaching the uppermost mapped layer. Once CO$_2$ arrived there, the detected footprint expanded rapidly within that layer. The resulting exponent therefore describes layer specific filling after delayed arrival, not a universal late time spreading regime.

Relative to the square-root reference, Aquistore and Weyburn--Midale fall on the subdiffusive side of the radius scaling, whereas the topmost Sleipner layer falls on the superdiffusive side. This terminology is used only as a descriptor of the fitted footprint exponent \(\beta\); it should not be read as evidence for a microscopic anomalous diffusion process.

The fitted exponents should therefore not be used to infer a unique value of the nonlinear transport exponent \(q\). The source free Barenblatt scaling gives \(1/4<\beta<1/2\) for \(0<q<1\), whereas the injection controlled capped regime gives the square-root reference. Aquistore and Weyburn--Midale are compatible with injection controlled growth, with site specific deviations caused by injection history and field geometry. Sleipner exceeds the square-root reference because the fitted time coordinate captures delayed arrival and rapid lateral filling of a particular stratigraphic layer. The comparison therefore supports a regime classification: near square-root injection controlled growth for Aquistore and Weyburn--Midale, and delayed layer filling for the topmost Sleipner plume layer.

\section{Conclusions}

We have developed a reduced plume model in which the primary scalar variable is the vertically averaged mobile  gas phase CO$_2$ inventory normalized by its maximum column value. This choice separates the mobile plume from the cumulative injected CO$_2$ and from CO$_2$ stored by dissolution, mineralization, residual trapping, or other retention mechanisms. The Global Buckley--Leverett balance provides the conservative structure of the reduction and identifies the net source/sink term, while the lateral spreading current is represented by an effective nonlinear coefficient \(D_u(u)\). The resulting equation is therefore a reduced plume scale transport law, not a microscopic constitutive model for relative permeability, capillary pressure, or compressibility.

The analytical construction gives two useful reference descriptions. In the tail only regime, the homogeneous nonlinear operator gives a Barenblatt-type compact support profile whose radius is controlled by both the nonlinear transport index \(q\) and the time dependence of the mobile quantity \(M_u(t)\). In the bounded regime, the physical constraint \(0\le u\le1\) produces a ful-thickness inner core joined to an outer Barenblatt-type tail. The corresponding radii \(a(t)\) and \(R(t)\) provide a simple criterion for the persistence or collapse of the core. Thus the model distinguishes between nonlinear redistribution of a mobile content and continued forcing by injection or removal by sinks.

Comparison with published time lapse seismic plume maps supports this regime interpretation. The fitted equivalent-radius exponents are \(\beta=0.449\) for Aquistore, \(\beta=0.478\) for Weyburn--Midale, and \(\beta=0.665\) for the topmost Sleipner layer. Aquistore and Weyburn--Midale lie close to the injection-controlled square-root reference, with Aquistore slightly below \(1/2\) and Weyburn--Midale nearly at \(1/2\). The topmost Sleipner layer gives a larger exponent, consistent with delayed arrival and rapid lateral filling of a stratigraphic layer rather than with a source free Barenblatt regime. The field data therefore do not determine a unique nonlinear exponent \(q\); instead, they separate near square-root injection controlled growth, layer filling behavior, and possible tail dominated redistribution.

The main limitation of the present analysis is that the source/sink term is represented through the mobile content history rather than through a fully resolved spatial model of injection, dissolution, trapping, and mineralization. The Barenblatt and capped profiles should therefore be read as analytical reference states for the reduced nonlinear operator, not as complete history matches of the monitored sites. Future extensions should incorporate measured injection schedules, spatially distributed sinks, and threshold dependent seismic detectability. Nonlocal or fractional generalizations may also be useful for testing whether anomalous plume footprint growth can be represented while preserving the bounded mobile quantity interpretation.

\section*{Data availability}
MATLAB script used to generate the numerical results  in this work, together with example input and output data, are openly available in the GitHub repository
\url{https://github.com/quantumfi/C02Plume}.

\section*{Acknowledgments}
The authors gratefully acknowledges Fatemeh Foroughirad and Morteza Nattagh-Najafi for valuable scientific discussion. F. Alonso-Marroquin also thanks Ruben Juanes for his hospitality at MIT and for taking the time to discuss the ideas that motivated this work.

\bibliographystyle{apsrev4-2}
\bibliography{plume,refs}
\clearpage

\end{document}